\batchmode
\makeatletter
\def\input@path{{"C:/Trabajo laptop/Mis articulos/Fusion 2021/GOSPA sensor management/"}}
\makeatother
\documentclass[british,english]{IEEEtran}
\usepackage[T1]{fontenc}
\usepackage[latin9]{inputenc}
\usepackage{amsmath}
\usepackage{amsthm}
\usepackage{amssymb}
\usepackage{stmaryrd}
\usepackage{graphicx}

\makeatletter
\theoremstyle{plain}

\theoremstyle{definition}
\newtheorem{defn}{\protect\definitionname}
\theoremstyle{plain}
\newtheorem{lem}{\protect\lemmaname}
\theoremstyle{definition}
\newtheorem{example}{\protect\examplename}

\usepackage{cite} 
\usepackage[margin=8pt,font=footnotesize]{caption}
\usepackage{algorithm}
\usepackage{algpseudocode}

\usepackage{amsmath}  
\allowdisplaybreaks

\makeatother

\usepackage{babel}
\addto\captionsbritish{\renewcommand{\definitionname}{Definition}}
\addto\captionsbritish{\renewcommand{\examplename}{Example}}
\addto\captionsbritish{\renewcommand{\lemmaname}{Lemma}}
\addto\captionsbritish{\renewcommand{\theoremname}{Theorem}}
\addto\captionsenglish{\renewcommand{\definitionname}{Definition}}
\addto\captionsenglish{\renewcommand{\examplename}{Example}}
\addto\captionsenglish{\renewcommand{\lemmaname}{Lemma}}
\addto\captionsenglish{\renewcommand{\theoremname}{Theorem}}
\providecommand{\definitionname}{Definition}
\providecommand{\examplename}{Example}
\providecommand{\lemmaname}{Lemma}
\providecommand{\theoremname}{Theorem}

\begin{document}
\title{An analysis on metric-driven multi-target sensor management: GOSPA
versus OSPA}
\author{Ángel F. García-Fernández\foreignlanguage{british}{$^{\star}${\normalsize{}$^{\circ}$}},
Marcel Hernandez\foreignlanguage{british}{$^{\star}$}, Simon Maskell\foreignlanguage{british}{$^{\star}$}\\
\foreignlanguage{british}{{\normalsize{}$^{\star}$}}{\normalsize{}Dept.
of Electrical Engineering and Electronics, University of Liverpool,
United Kingdom}\\
\foreignlanguage{british}{{\normalsize{}$^{\circ}$}}{\normalsize{}ARIES
Research Centre, Universidad Antonio de Nebrija, Spain}\\
{\normalsize{}Emails: \{angel.garcia-fernandez, marcel.hernandez,
s.maskell\}@liverpool.ac.uk}\thanks{This work was supported by the UK Ministry of Defence under contract number DSTLX-1000143908.}}

\maketitle
\thispagestyle{empty}
\begin{abstract}
This paper presents an analysis on sensor management using a cost
function based on a multi-target metric, in particular, the optimal
subpattern-assignment (OSPA) metric, the unnormalised OSPA (UOSPA)
metric and the generalised OSPA (GOSPA) metric ($\alpha=2$). We consider
the problem of managing an array of sensors, where each sensor is
able to observe a region of the surveillance area, not covered by
other sensors, with a given sensing cost. We look at the case in which
there are far-away, independent potential targets, at maximum one
per sensor region. In this set-up, the optimal action using GOSPA
is taken for each sensor independently, as we may expect. On the contrary,
as a consequence of the spooky effect at a distance in optimal OSPA/UOSPA
estimation, the optimal actions for different sensors using OSPA and
UOSPA are entangled.
\end{abstract}


\section{Introduction}

Surveillance in a large area is often carried out by multiple sensors
that have different fields of view and possibly different sensing
capabilities \cite{Kreucher07}. In order to maximise sensing resources,
sensors may be equipped with different sensing modes and may be able
to move to observe different regions. The objective of sensor management
is therefore to decide the actions of the sensors (e.g., sensing mode
or movement) to maximise surveillance performance \cite{Hero_book08}.
How to measure surveillance performance depends on the task at hand
and may comprise different aspects, such as tracking accuracy, the
number of false targets and detection performance. 

Sensor management problems in dynamic systems are usually posed as
partially observed Markov decision processes (POMDPs) \cite{Hero_book08,Thrun_book05}.
In this setting, the aim is to select the action that minimises the
expected value of a cost function (or maximise an expected reward)
in a certain time window. Myopic sensor management refers to making
the best decision only considering the current time step, without
looking at possible costs in the future. This approach can be improved,
with an increase in computational complexity, by non-myopic sensor
management, which considers the expected cost for each action across
several future time steps.

In surveillance, targets may appear, move and disappear in the area
of interest \cite{Mahler_book14}, and sensor management has to balance
decisions regarding exploration, exploitation, and sensing costs \cite{Kreucher07}.
That is, shall we use resources to track the already detected targets
better or explore new areas in search of new targets? From a POMDP
perspective, the solution to this problem requires the posterior density
of the set of targets for each sensing action, as it contains all
information over detected and undetected targets. For example, these
two types of information are explicit in the Poisson multi-Bernoulli
mixture filter \cite{Williams15b,Angel18_b,Bostrom-Rost21_early}.
We also require a suitable cost function that is able to measure performance
and therefore drives the sensor actions in an appropriate way.

In the literature, there are different types of cost functions. For
example, the posterior Cramér-Rao lower bound (PCRLB) \cite{Hernandez_inbook13,Tharmarasa07,Bell15}
is a bound on the mean square error, so a cost function based on it
must be extended with complementary criteria to deal with an unknown
and variable number of targets. Information-theoretic approaches maximise
the expected gain in information of the posterior w.r.t. the predicted
density \cite{Kreucher07,Aoki11,Ristic11b,Beard17b}. While these
approaches can work well, they have the drawback that it is not very
clear what maximising the information gain means in practice.

Multiple target filtering performance is usually evaluated via metrics
on sets of targets. Therefore, metric-driven sensor management, in
which the cost function is related to a metric, provides a clear interpretation
of what the objective is, and also measurable results in terms of
performance evaluation. In this respect, the optimal subpattern assignment
metric (OSPA) is a widely-used metric \cite{Schuhmacher08_b,Schuhmacher08}.
A sensor management algorithm based on the OSPA metric was proposed
in \cite{Gostar15}. 

However, the OSPA metric does not penalise the main errors of interest
in most multi-target estimation tasks: localisation error for properly
detected targets, and number of missed and false targets \cite{Drummond92}.
A metric designed to penalise these errors is the generalised OSPA
(GOSPA) metric \cite{Rahmathullah17}, which also avoids the spooky
effect in optimal estimation using OSPA \cite{Angel19_d}. An algorithm
for GOSPA-based sensor management  with one potential target is provided
in \cite[Chap. 6]{Ubeda_thesis18}.

This paper provides a theoretical analysis on metric-driven (myopic)
sensor management, in particular comparing OSPA, unnormalised OSPA
and the GOSPA metrics. We tackle a problem in which we manage a collection
of sensors, each of which should decide whether or not to measure
a single potential target, which is independent and far away from
the rest of the targets, with a certain sensing cost.   The results
show that the GOSPA metric provides optimal actions that are separable.
That is, each sensor makes its own decisions, which is the expected
result in this type of problem. On the contrary, the optimal actions
for the OSPA and UOSPA metrics are entangled in the sense that the
optimal action of a sensor depends on far-away independent potential
targets, outside its field of view. The underlying reason for this
entanglement is the spooky effect at a distance in optimal OSPA and
UOSPA estimation, in which the optimal estimation of a potential target
is affected by far away, independent potential targets. 

The rest of the paper is organised as follows. Section \ref{sec:Problem-formulation}
presents the problem formulation. Section \ref{sec:Analysis-I} presents
an analysis with one potential target, and Section \ref{sec:Analysis-II}
with $N$ potential targets. Conclusions are drawn in Section \ref{sec:Conclusion}.

\section{Problem formulation\label{sec:Problem-formulation}}

We tackle the problem of (myopic) sensor management in multi-target
systems by using a cost function based on a multi-target metric.

\subsection{Metrics}

Let $c$ and $p$ be two real numbers such that $c>0$ and $1\leq p<\infty$.
We use $d\left(\cdot,\cdot\right)$ to denote a metric on the single
target space, which is typically $\mathbb{R}^{n_{x}}$, and $d^{\left(c\right)}\left(\cdot,\cdot\right)=\min\left(d\left(\cdot,\cdot\right),c\right)$.
The set of all permutations of $\left\{ 1,...,n\right\} $ where $n\in\mathbb{N}$
is $\prod_{n}$. Any element $\pi\in\prod_{n}$ is written as $\pi=\left(\pi\left(1\right),..,\pi\left(n\right)\right)$.
Also, let $X=\left\{ x_{1},...,x_{\left|X\right|}\right\} $ and $Y=\left\{ y_{1},...,y_{\left|Y\right|}\right\} $
denote two finite sets of targets, with $\left|X\right|\leq\left|Y\right|$,
and $\left|X\right|$ being the cardinality (number of elements) of
set $X$.
\begin{defn}
The OSPA metric between $X$ and $Y$ for $\left|Y\right|>0$ is \cite{Schuhmacher08_b,Schuhmacher08}
\begin{align*}
 & d_{p}^{\left(c\right)}\left(X,Y\right)\\
 & =\min_{\pi\in\prod_{\left|Y\right|}}\left(\frac{1}{\left|Y\right|}\sum_{i=1}^{\left|X\right|}d^{\left(c\right)}\left(x_{i},y_{\pi\left(i\right)}\right)^{p}+c^{p}\left(\left|Y\right|-\left|X\right|\right)\right)^{1/p}.
\end{align*}
For $\left|Y\right|=0$ and $\left|X\right|=0$, $d_{p}^{\left(c\right)}\left(\emptyset,\emptyset\right)=0$.
$\boxempty$
\end{defn}
The UOSPA metric corresponds to OSPA without the division by $\left|Y\right|$
and is proportional to the metric in \cite{Barrios17}.

Compared to the OSPA/UOSPA metrics, the GOSPA metric has an additional
parameter $\alpha$ that controls the cardinality mismatch penalty.
Importantly, only for $\alpha=2$, the GOSPA metric can be written
in terms of costs corresponding to localisation errors for properly
detected targets, missed and false targets, which are usually the
penalties of interest in multiple target estimation. 

Let $\gamma$ be an assignment set between $\left\{ 1,...,\left|X\right|\right\} $
and $\left\{ 1,...,\left|Y\right|\right\} $, which meets $\gamma\subseteq\left\{ 1,...,\left|X\right|\right\} \times\left\{ 1,...,\left|Y\right|\right\} $,
$\left(i,j\right),\left(i,j'\right)\in\gamma\rightarrow j=j'$, and
$\left(i,j\right),\left(i',j\right)\in\gamma\rightarrow i=i'$. The
last two properties ensure that every $i$ and $j$ gets at most one
assignment. We denote the set of all possible $\gamma$ as $\Gamma$.
\begin{defn}
\label{def:GOSPA_alpha2}The GOSPA metric ($\alpha=2$) between $X$
and $Y$ is \cite[Prop. 1]{Rahmathullah17}
\begin{align}
 & d_{p}^{\left(c,2\right)}\left(X,Y\right)\nonumber \\
 & =\min_{\gamma\in\Gamma}\left(\sum_{\left(i,j\right)\in\gamma}d^{p}\left(x_{i},y_{j}\right)+\frac{c^{p}}{2}\left(\left|X\right|+\left|Y\right|-2\left|\gamma\right|\right)\right)^{1/p}.\boxempty\label{eq:GOSPA_alpha2}
\end{align}
\end{defn}
The first term in (\ref{eq:GOSPA_alpha2}) represents the localisation
errors (to the $p$-th power) for assigned targets (properly detected
ones), which meet $\left(i,j\right)\in\gamma$. The terms $\frac{c^{p}}{2}\left(\left|X\right|-\left|\gamma\right|\right)$
and $\frac{c^{p}}{2}\left(\left|Y\right|-\left|\gamma\right|\right)$
represent the costs (to the $p$-th power) for missed and false targets.

\subsection{Metric-driven sensor management}

As we focus on myopic sensor management, which only involves a single
time step, we do not include the time index in the notation for simplicity.
At the current time step, before observing the current measurements,
all information of interest about the current set $X$ of targets
is contained in its predicted density $p\left(\cdot\right)$, which
represents the density of the current state given past measurements. 

In the sensor management problem we consider, at each time step, we
must take an action $a\in\mathbb{A}$, where $\mathbb{A}$ is the
possible sets of actions, that affects the way we observe the current
set of targets. Mathematically, we can write that the conditional
density of the set $Z$ of measurements at the current time step depends
on $X$ and $a$, such that $p\left(Z|X;a\right)$. Given an action
$a$ and an observation $Z$, the posterior is \cite{Mahler_book14}
\begin{align}
p\left(X|Z;a\right) & =\frac{p\left(Z|X;a\right)p\left(X\right)}{p\left(Z;a\right)}\label{eq:posterior}
\end{align}
where the normalising constant is
\begin{align}
p\left(Z;a\right) & =\int p\left(Z|X;a\right)p\left(X\right)\delta X.\label{eq:density_measurement}
\end{align}

The objective is to determine the best action $a$ by minimising a
cost function involving the current time step. Multiple target tracking
algorithms are usually evaluated via metrics, so in order to improve
their performance in this sense, the cost function should include
the mean square error, or a variant \cite[Prop. 2]{Rahmathullah17},
for each action.

Given the posterior (\ref{eq:posterior}), we can obtain a multi-target
estimate $\hat{X}\left(Z,a\right)$. The resulting mean square GOSPA
(MSGOSPA) error given $Z$ and action $a$ is
\begin{align}
 & \mathrm{E}\left[\left(d_{2}^{\left(c,2\right)}\left(X,\hat{X}\left(Z,a\right)\right)\right)^{2}|Z;a\right]\nonumber \\
 & =\int\left(d_{2}^{\left(c,2\right)}\left(X,\hat{X}\left(Z,a\right)\right)\right)^{2}p\left(X|Z;a\right)\delta X.\label{eq:posterior_MSGOSPA}
\end{align}
The lowest MSGOSPA error is achieved if we select $\hat{X}\left(Z,a\right)$
to be the estimator that minimises (\ref{eq:posterior_MSGOSPA}).
That is, if $\hat{X}\left(Z,a\right)$ is the optimal minimum MSGOSPA
(MMSGOSPA) error estimator. The resulting MMSGOSPA error, given $Z$,
is
\begin{align}
 & \mathrm{MMSGOSPA}\left(Z,a\right)\nonumber \\
 & \;=\min_{\hat{X}\left(Z,a\right)}\mathrm{E}\left[\left(d_{2}^{\left(c,2\right)}\left(X,\hat{X}\left(Z,a\right)\right)\right)^{2}|Z;a\right].\label{eq:MMSGOSPA}
\end{align}

The action that minimises the MMSGOSPA error, averaged over all $Z$,
meets
\begin{align}
 & \underset{a}{\arg\min}\int\mathrm{MMSGOSPA}\left(Z,a\right)p\left(Z;a\right)\delta Z.
\end{align}

In practice, different actions may have different sensing costs, e.g.,
associated to energy consumption, so we can add a sensing cost that
penalises each action. Combining the above results with an additive
sensor cost, the optimal action is then calculated as in the following
lemma.
\begin{lem}
\label{lem:Optimal_action}The action $\hat{a}$ that minimises the
mean square GOSPA error with an additive sensing cost $s\left(a\right)$
meets
\begin{align}
\hat{a} & =\underset{a}{\arg\min}\,c\left(a\right)\label{eq:lem_optimal_action}\\
c\left(a\right) & =\int\mathrm{MMSGOSPA}\left(Z,a\right)p\left(Z;a\right)\delta Z+s\left(a\right)\label{eq:c_action}
\end{align}
where $\mathrm{MMSGOSPA}\left(Z,a\right)$ is the minimum mean square
GOSPA error conditioned on $Z$, which is given by (\ref{eq:MMSGOSPA}). 
\end{lem}
The result in Lemma \ref{lem:Optimal_action} directly extends to
OSPA and UOSPA by using the corresponding metric. In most cases, it
is intractable to obtain the optimal action $\hat{a}$, as even the
optimal estimator, which is required to obtain the MMSGOSPA is intractable.
Therefore, in practice, we have to develop algorithms to approximate
(\ref{eq:lem_optimal_action}). In this paper however, we analyse
two cases in which the optimisation (\ref{eq:lem_optimal_action})
is closed-form, to be able to draw conclusions without approximations.

\section{Analysis I: One potential target\label{sec:Analysis-I}}

We analyse the problem of deciding whether we should measure or not
a single potential target. We consider that the prior is a Bernoulli
density of the form
\begin{align}
p\left(X\right) & =\begin{cases}
1-r & X=\emptyset\\
r\delta\left(x-\overline{x}\right) & X=\left\{ x\right\} \\
0 & \mathrm{otherwise}
\end{cases}\label{eq:Bernoulli_density}
\end{align}
where $r\in\left[0,1\right]$ is the probability of existence, $\delta\left(\cdot\right)$
is a Dirac delta and $\overline{x}$ is the mean location if the target
exist. It should be noted that we consider that the location of the
target is known to obtain closed-form expressions. The same conclusions
of this analysis hold if the single-target density is a Gaussian with
a small variance, compared to $c$.

The first action, $a=0$, represents that we do not observe the area
where the target lies. This can be represented mathematically by setting
$p\left(Z;0\right)=\delta_{\emptyset}\left(Z\right)$, which represents
a multi-target Dirac delta centered at $\emptyset$ \cite{Mahler_book14}
and indicates that we measure $Z=\emptyset$ with probability one.

The second action, $a=1$, represents that we observe the area where
the target lies. In the standard measurement model \cite{Mahler_book14},
observing an area does not directly imply that the target is observed.
In our analysis, under action $a=1$, we assume 
\begin{itemize}
\item A1 A target with state $x$ may be detected with a probability $p^{D}$
and generates a measurement $z$ with density $l\left(z|x\right)$,
and may be misdetected with probability $1-p^{D}$.
\item A2 There is no clutter.
\end{itemize}
We consider that clutter in non-existent for tractability. We also
consider that the sensing costs are $s\left(0\right)=0$, and $s\left(1\right)=s$
where $s\geq0$. 

We prove in Appendix \ref{sec:AppendixA} that the cost $c\left(a\right)$,
with the GOSPA metric, for actions 0 and 1 is
\begin{align}
c\left(0\right) & =\frac{c^{2}}{2}t\left(r\right)\label{eq:c0_example1}\\
c\left(1\right) & =\frac{c^{2}}{2}t\left(\frac{r\left(1-p^{D}\right)}{1-rp^{D}}\right)\left(1-rp^{D}\right)+s\label{eq:c1_example1}
\end{align}
where
\begin{align}
t\left(r\right) & =\begin{cases}
r & r<0.5\\
1-r & r\geq0.5.
\end{cases}\label{eq:t_function}
\end{align}

It should be noted that the costs do not depend on $l\left(z|x\right)$.
The critical part of the measurement model is $p^{D}$. The costs
$c\left(0\right)$ and $c\left(1\right)$ for the OSPA and UOSPA metrics
are also (\ref{eq:c0_example1}) and (\ref{eq:c1_example1}) but using
$c^{2}$ instead of $\frac{c^{2}}{2}$. Therefore, the three metrics
behave analogously to solve this sensor management problem and we
focus on the GOSPA case. Major differences between the metrics will
arise when we consider two potential targets, as will be analysed
in Section \ref{sec:Analysis-II}.

We can compute the optimal action, which is given by (\ref{eq:lem_optimal_action}),
in closed form. As proved in Appendix \ref{sec:AppendixA}, for $s>0$,
we measure the target (action $\hat{a}=1$) if the probability $r$
of existence and the sensing cost $s$ meet
\begin{align}
\frac{2s}{c^{2}p^{D}}< & r<\frac{\frac{c^{2}}{2}-s}{c^{2}\left(1-\frac{1}{2}p^{D}\right)}\label{eq:optimal_action_analysisI_1}\\
s & <\frac{c^{2}p^{D}}{4}.\label{eq:optimal_action_analysisI_2}
\end{align}
If (\ref{eq:optimal_action_analysisI_1}) and (\ref{eq:optimal_action_analysisI_2})
are not met, the optimal action is not to measure the target $\hat{a}=0$.
For $s=0$, cost $c\left(1\right)$ is never higher than $c\left(0\right)$
so one can always choose $\hat{a}=1$. It is interesting to note that
for sufficiently low sensing costs, according to (\ref{eq:optimal_action_analysisI_2}),
we only take action if the probability of existence is not too low
or too high. That is, we measure the target if we are not very confident
whether the target exists or not, and we therefore gain valuable information
by measuring it.

\begin{figure}
\begin{centering}
\includegraphics[scale=0.6]{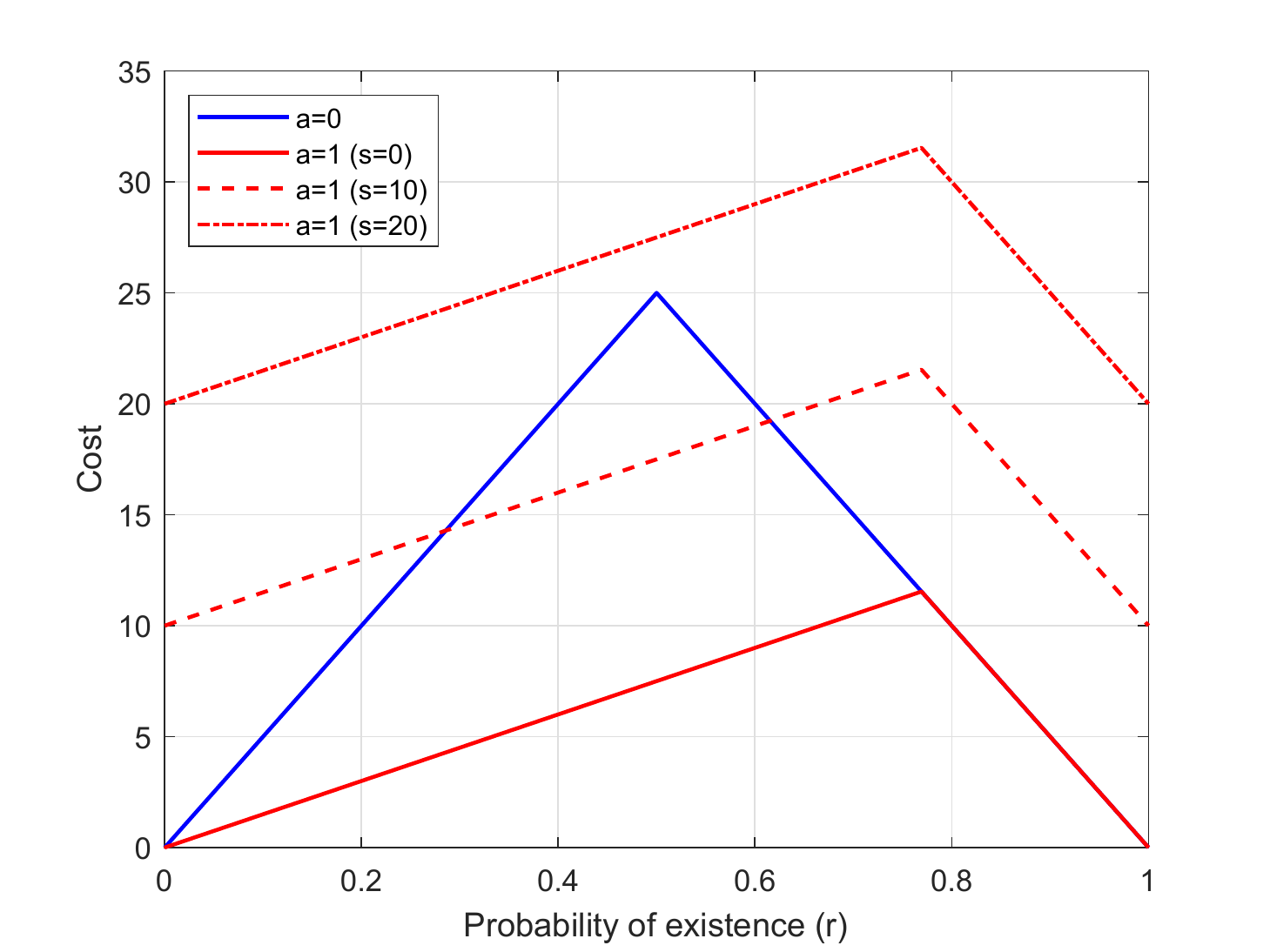}
\par\end{centering}
\caption{\label{fig:Cost-varying_s}Cost $c\left(a\right)$ (for GOSPA) as
a function of the probability $r$ of existence for $a=0$ and $a=1$,
$p^{D}=0.7$, and $s\in\left\{ 0,10,20\right\} $. }
\end{figure}

\begin{figure}
\begin{centering}
\includegraphics[scale=0.3]{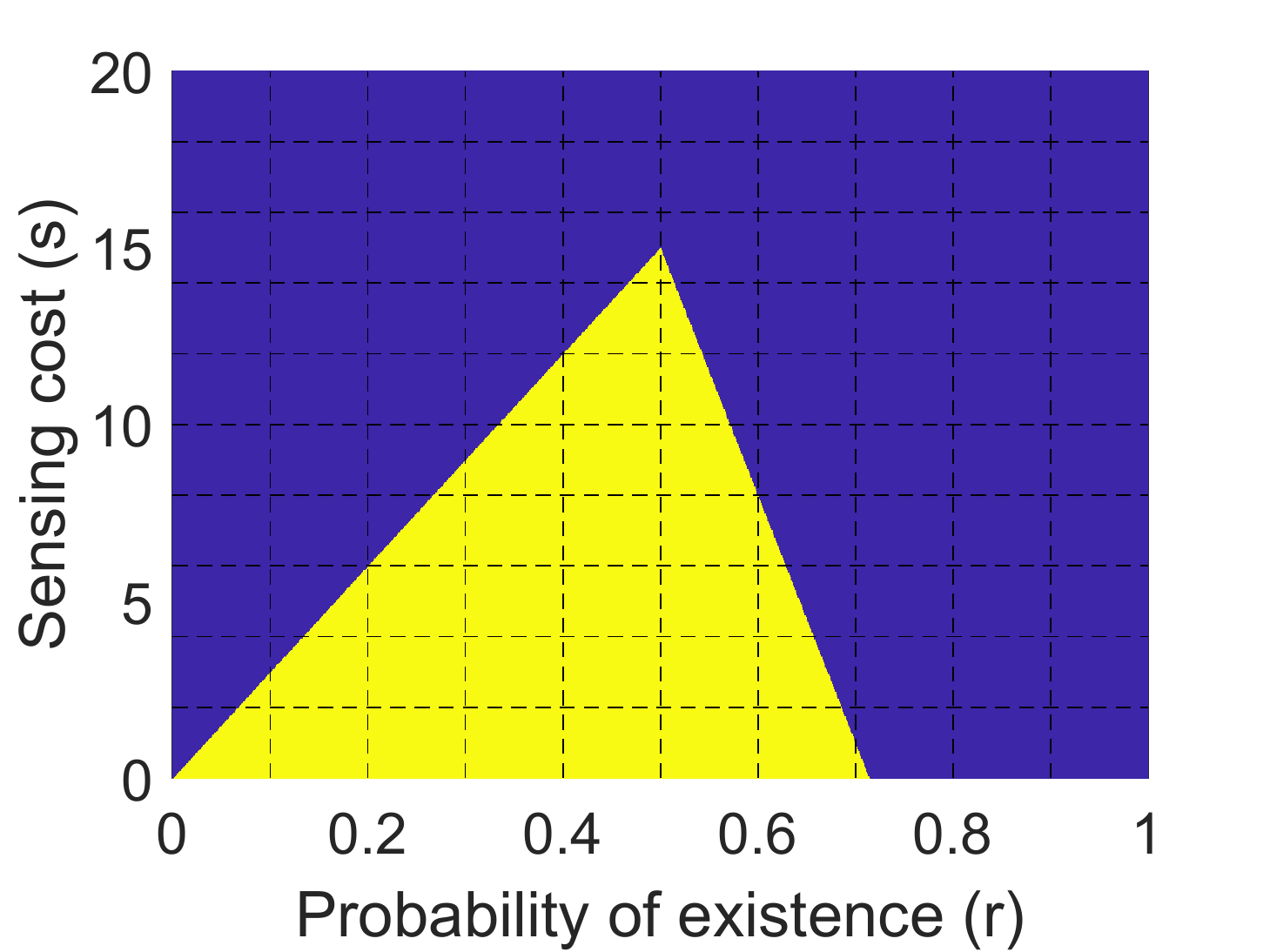}\includegraphics[scale=0.3]{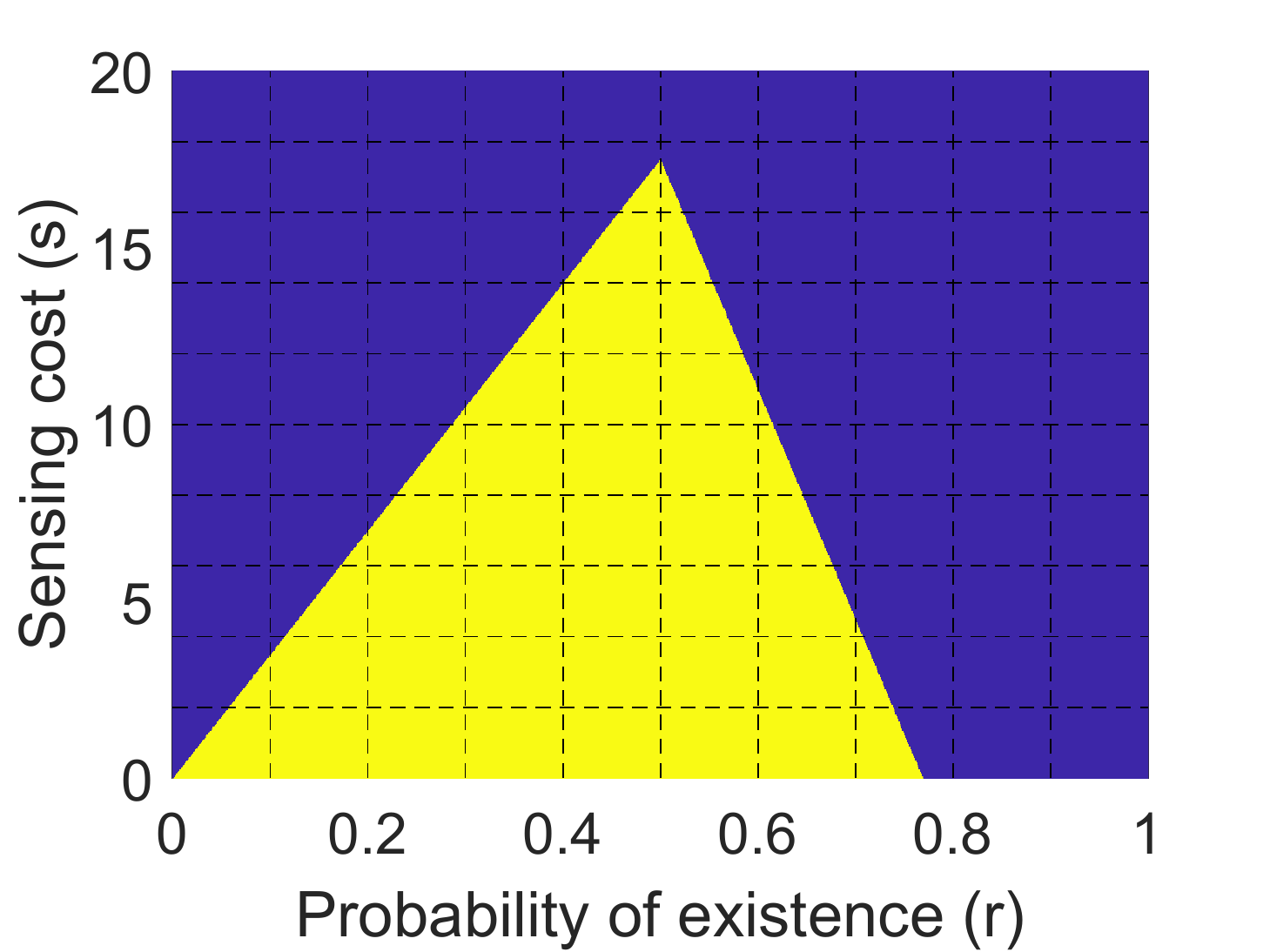}
\par\end{centering}
\caption{\label{fig:Optimal-action-1target_GOSPA}Optimal action (for GOSPA)
as a function of the probability $r$ of existence and sensing cost
$s$ for $p^{D}=0.6$ (left) and $p^{D}=0.7$ (right). Action $a=0$
is shown in blue and $a=1$ in yellow. As $p^{D}$ decreases, the
size of the area to take action $1$ decreases because of a reduction
in the value of making an observation. }

\end{figure}

We proceed to illustrate these results in two examples.
\begin{example}
This example analyses the effect of varying $s$ and $r$ to select
the optimal action $\hat{a}$. We use $c=10$, $p^{D}=0.7$ and $s\in\left\{ 0,10,20\right\} $.
The resulting costs $c\left(a\right)$ against $r$ are shown in Figure
\ref{fig:Cost-varying_s}. We should first recall that for a given
$r$, the best action $\hat{a}$ is the one with lowest cost. For
$s=0$, we have that for $r$ less than 0.77, the optimal action is
$\hat{a}=1$. For $r$ higher than 0.77, both actions have equal cost.
The explanation is that, once $r$ is high enough, the optimal estimation
for both actions and when we receive no measurement or a target detection,
is always to estimate $\hat{X}\left(Z,a\right)=\left\{ \overline{x}\right\} $,
and therefore, using (\ref{eq:c_action}), we have $c\left(0\right)=c\left(1\right)$.

If $s$ increases, the curve for $c\left(1\right)$ is shifted upwards
along the $y$-axis. For $s=20$, choosing $a=1$ becomes too costly
and it is best not to measure the target for any $r$. For $s=10$,
we measure the target if the probability of existence is within a
certain interval, given by (\ref{eq:optimal_action_analysisI_1}).
The optimal action against $r$ and $s$ is shown in Figure \ref{fig:Optimal-action-1target_GOSPA}.
$\diamondsuit$
\end{example}

\begin{figure}
\begin{centering}
\includegraphics[scale=0.6]{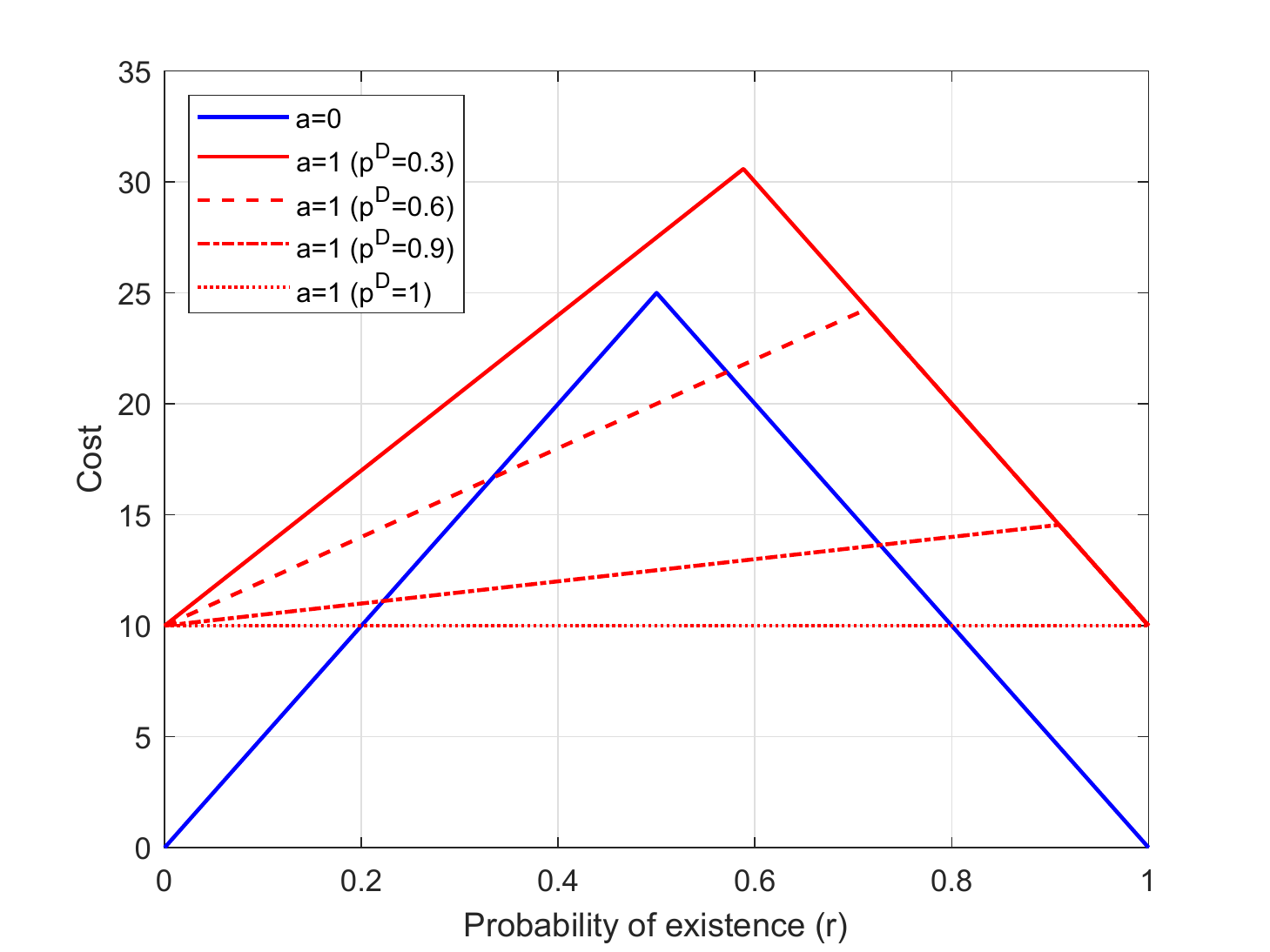}
\par\end{centering}
\caption{\label{fig:Cost-varying-pd}Cost $c\left(a\right)$ (for GOSPA) as
a function of the probability $r$ of existence for $a=0$ and $a=1$,
$s=10$, and $p^{D}\in\left\{ 0.3,0.6,0.9,1\right\} $. }
\end{figure}

\begin{example}
This example analyses the effect of varying $p_{D}$ and $r$ to select
the optimal action $\hat{a}$. We set $s=10$ and $p^{D}\in\left\{ 0.3,0.6,0.9,1\right\} $
and the resulting costs against $r$ are shown in Figure \ref{fig:Cost-varying-pd}.
For $p^{D}=1$, $c\left(1\right)$ is a constant equal to $s$, as
there is no estimation error. Even for $p_{D}=1$, it is sometimes
best not to take action 0 and not measure the target, if $r$ is sufficiently
low or sufficiently high. As $p^{D}$ decreases, the slope of the
first segment of $c\left(1\right)$ increases, and the range of values
of $r$ in which it is best to take action $1$ decreases. For $p^{D}=0.3$,
the best action is not to measure for all values of $r$. That is,
the information we expect to obtain from measuring is not sufficient
to compensate the sensing cost $s$. $\diamondsuit$
\end{example}

\section{Analysis II: $N$ potential targets\label{sec:Analysis-II}}

We analyse the problem of deciding whether we should measure or not
$N$ potential targets far away from each other. The prior is multi-Bernoulli
with known target locations such that \cite{Mahler_book14}
\begin{align}
p\left(X\right) & =\sum_{X_{1}\uplus...\uplus X_{N}=X}\prod_{i=1}^{N}f_{i}\left(X_{i}\right),\label{eq:MB_density}\\
f_{i}\left(X\right) & =\begin{cases}
1-r_{i} & X=\emptyset\\
r_{i}\delta\left(x-\overline{x}_{i}\right) & X=\left\{ x\right\} \\
0 & \mathrm{otherwise}
\end{cases}
\end{align}
where $r_{i}$ is the probability of existence of the $i$-th Bernoulli
and $\overline{x}_{i}$ is the location of the $i$-th Bernoulli component.
All Bernoulli components are far from each other $d\left(\overline{x}_{i},\overline{x}_{j}\right)>c$
for $i\neq j$.

An action is represented as $a=\left[a_{1},...,a_{N}\right]$ where
$a_{i}\in\left\{ 0,1\right\} $ with $a_{i}=0$ if we do not observe
the $i$-th Bernoulli and $a_{i}=1$ if we observe it. We assume an
additive sensor cost
\begin{align}
s\left(a\right) & =s\sum_{i=1}^{N}a_{i}
\end{align}
where $s$ is the cost of activating a single sensor. 

In this analysis, under action $a=1$, we assume A1, A2 and
\begin{itemize}
\item A3 If $l\left(z|\overline{x}_{i}\right)>0$ then $l\left(z|\overline{x}_{j}\right)=0$
for $i\neq j.$
\end{itemize}
A3 implies that targets do not generate measurements in the areas
of other targets. 

We prove in Appendix \ref{sec:AppendixB} that the cost $c\left(a\right)$,
with the GOSPA metric, is
\begin{align}
c\left(a\right) & =\sum_{i=1}^{N}c_{i}\left(a_{i}\right)\label{eq:cost_GOSPA_II_a}\\
c_{i}\left(0\right) & =\frac{c^{2}}{2}t\left(r_{i}\right)\\
c_{i}\left(1\right) & =\frac{c^{2}}{2}t\left(\frac{r_{i}\left(1-p^{D}\right)}{1-r_{i}p^{D}}\right)\left(1-r_{i}p^{D}\right)+s.\label{eq:cost_GOSPA_II_c}
\end{align}
The cost for GOSPA is additive over the Bernoulli components, which
implies that the optimal actions are computed independently for each
Bernoulli using (\ref{eq:optimal_action_analysisI_1}) and (\ref{eq:optimal_action_analysisI_2}).

For OSPA and UOSPA, there is not such a simplification. One should
use Lemma \ref{lem:Optimal_action} by evaluating all possible estimates
and actions. In Appendix \ref{sec:AppendixB}, we provide more simplified
expressions for 2 Bernoulli components, which are used for the following
illustrative example.

\begin{figure}
\begin{centering}
\includegraphics[scale=0.6]{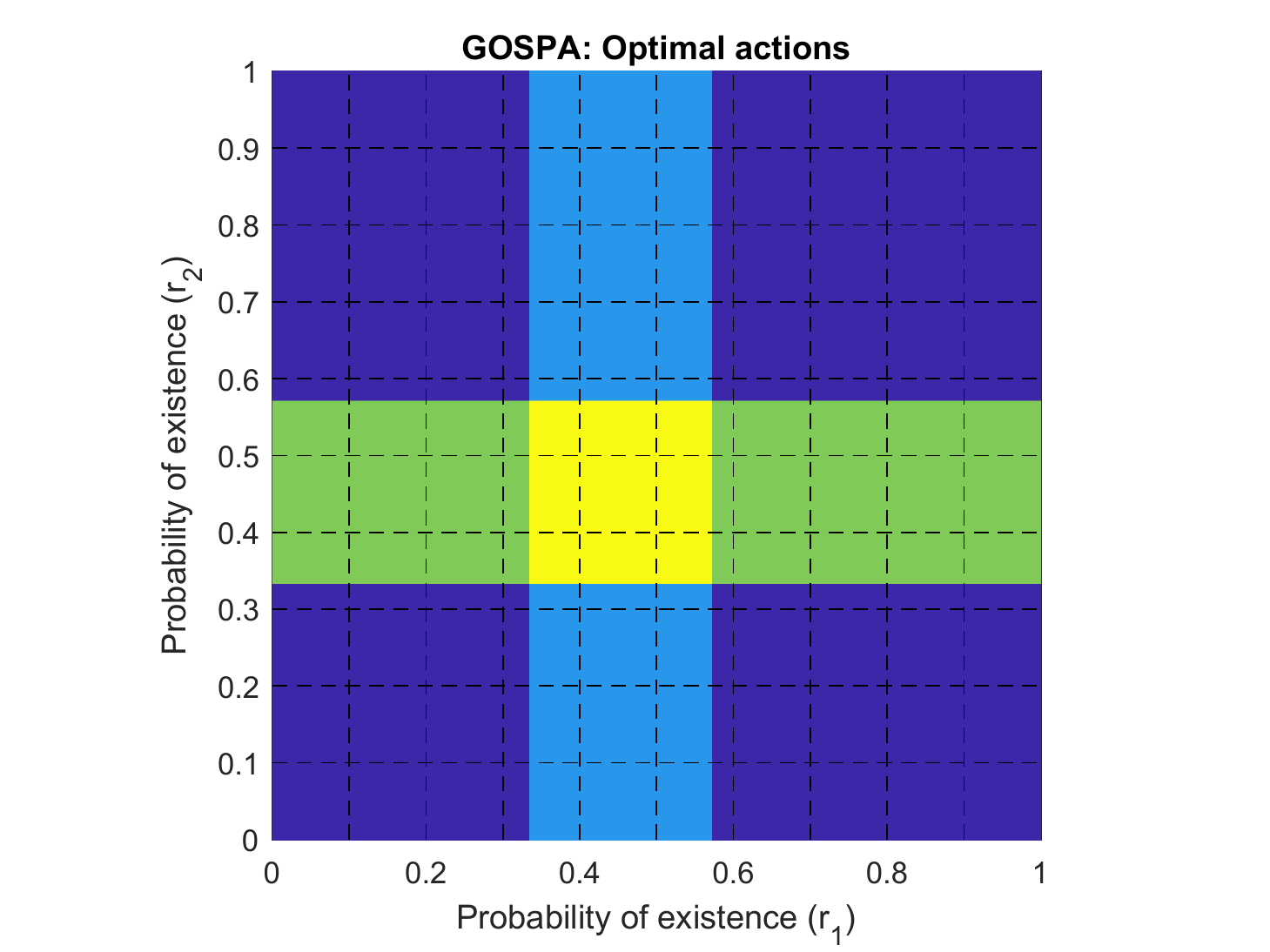}
\par\end{centering}
\begin{centering}
\includegraphics[scale=0.6]{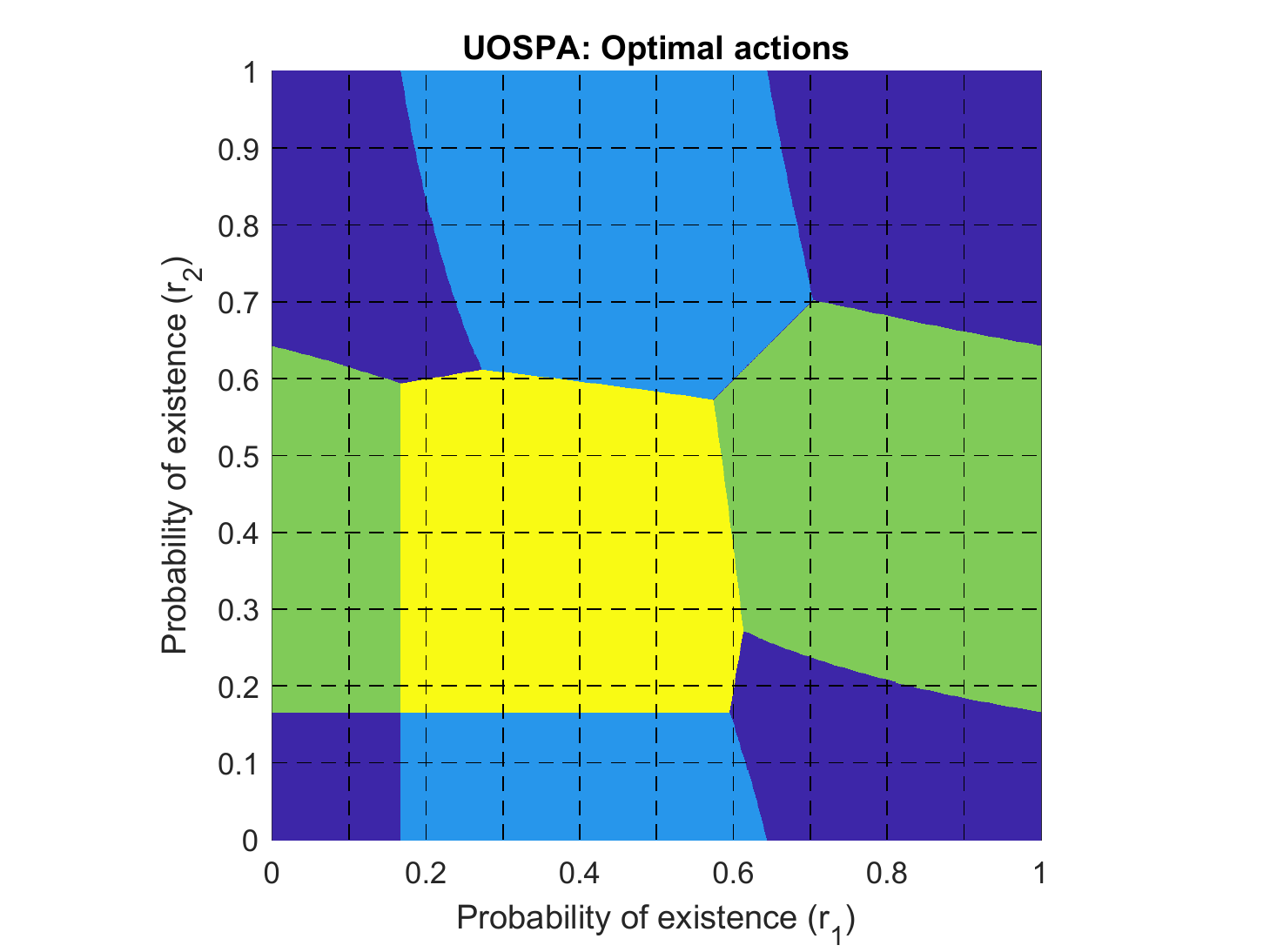}
\par\end{centering}
\begin{centering}
\includegraphics[scale=0.6]{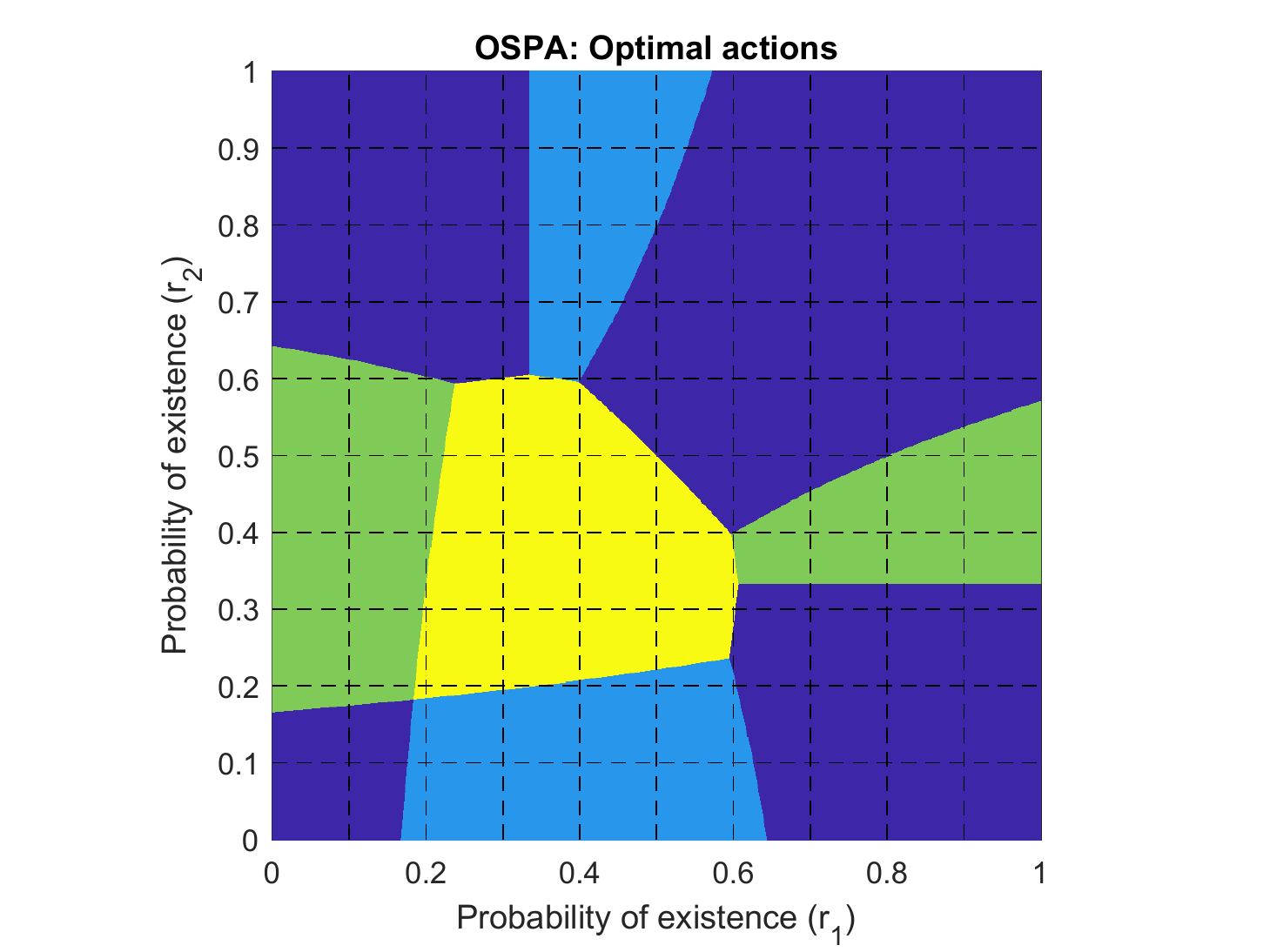}
\par\end{centering}
\caption{\label{fig:Optimal-actions-2-Bern}Optimal actions for GOSPA, UOSPA
and OSPA metrics against the existence probabilities of two Bernoulli
components. The colours for the optimal actions are: dark blue $a=\left(0,0\right)$,
light blue $a=\left(1,0\right)$, green $a=\left(0,1\right)$ and
yellow $a=\left(1,1\right)$. For GOSPA, the decision of whether to
measure a target or not is taken independently of the other. For UOSPA
and OSPA, there is coupling in the optimal actions. }
\end{figure}

\begin{example}
This example analyses the optimal actions for $N=2$ Bernoulli components
against the probability of existences. We use $c=10$, $p^{D}=0.6$
and $s=10$. The resulting optimal actions against $r_{1}$ and $r_{2}$
are shown in Figure \ref{fig:Optimal-actions-2-Bern}. The optimal
action is independent for each target using GOSPA. This is what is
expected in most sensor systems observing non-overlapping, distant
regions and whose sensor cost is additive. Note that the costs for
each action for each target are similar to the ones in Figure \ref{fig:Cost-varying-pd}.
That is, we measure a given target if its probability of existence
is not too low or too high, with optimal action indicated by (\ref{eq:optimal_action_analysisI_1})
and (\ref{eq:optimal_action_analysisI_2}). 

When we use UOSPA and OSPA, there is influence of far-away independent
targets on the optimal sensing mode for a sensor covering a different
surveillance area. That is, for a given $r_{1}$, the sensing action
$a_{1}$ for this target depends on $r_{2}$, the probability of existence
of the other target. The choice of the optimal action is clearly counterintuitive
for some choices of $r_{1}$ and $r_{2}$. For example, fixing $r_{2}=0.6$,
the optimal action as a function of $r_{1}$ for both OSPA and UOSPA
is shown in Figure \ref{fig:Optimal-action-fixed_r2}. For OSPA, starting
with $r_{1}=0$, we measure target 2. Then, as $r_{1}$ increases,
we stop measuring target 2, even though nothing has changed related
to target 2. If $r_{1}$ keeps increasing, we measure both targets.
Then, we first stop measuring target 2, and then we stop measuring
both targets.  $\diamondsuit$

\begin{figure}
\begin{centering}
\includegraphics[scale=0.3]{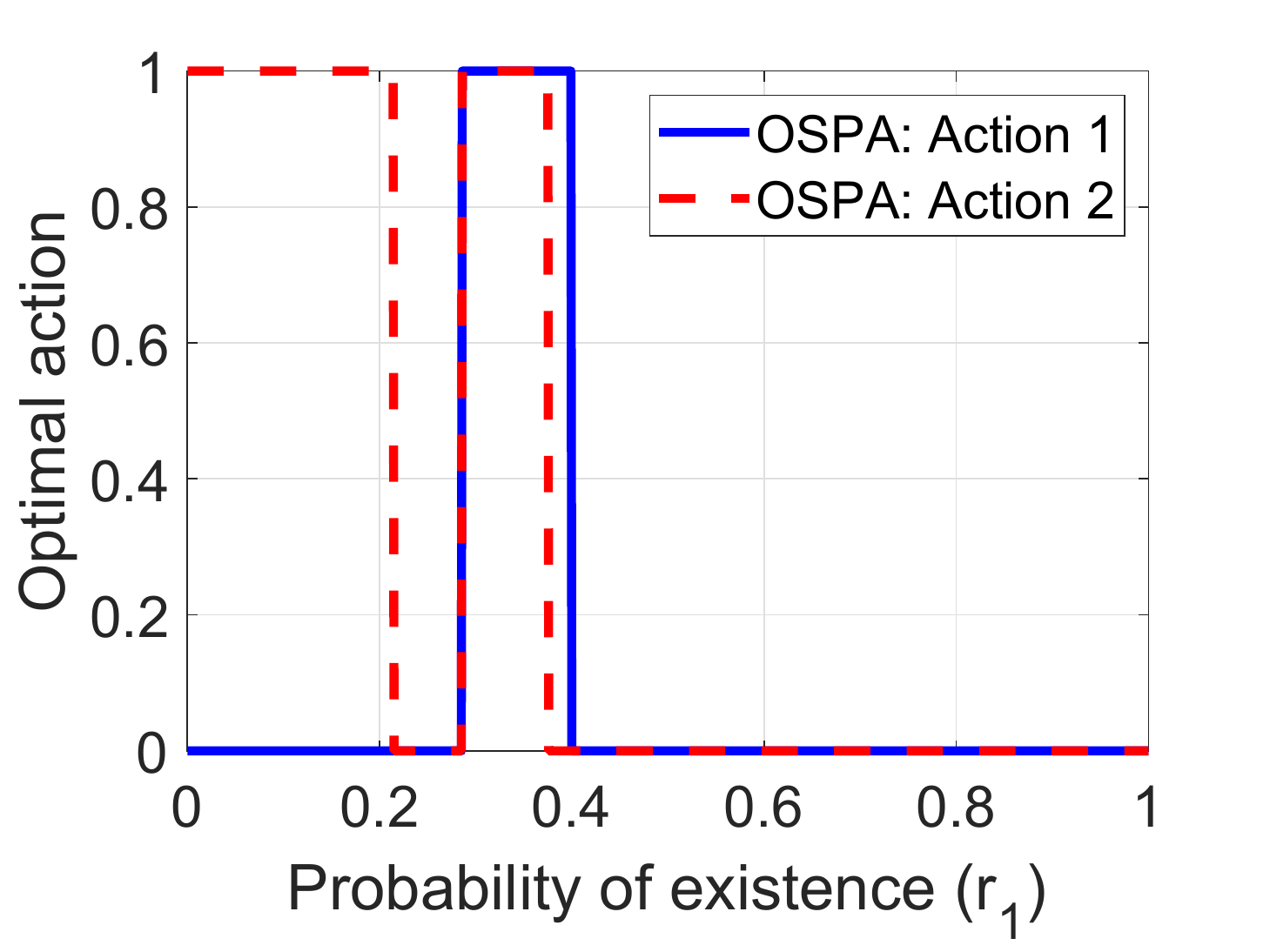}\includegraphics[scale=0.3]{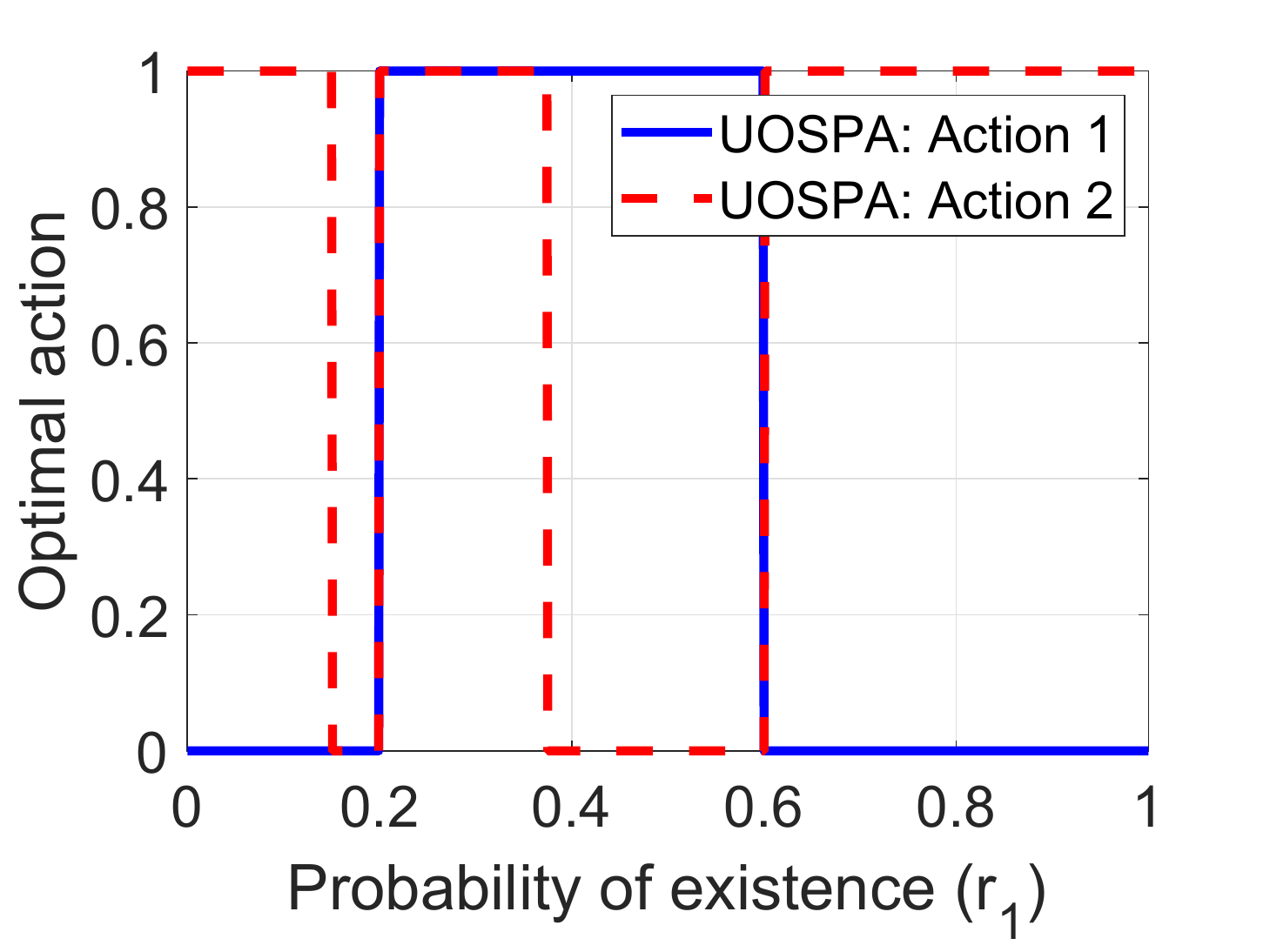}
\par\end{centering}
\caption{\label{fig:Optimal-action-fixed_r2}Optimal action as a function of
$r_{1}$ fixing $r_{2}=0.6$ for OSPA and UOSPA. As opposed to GOSPA,
a change in $r_{1}$ modifies the optimal action for both sensors.}

\end{figure}

The explanation of this behaviour of the optimal actions is that in
order to minimise a metric-driven sensor management problem, we must
consider the optimal estimators, as these minimise the corresponding
error. UOSPA and OSPA optimal estimators suffer from spooky effect
at a distance, in which the optimal estimation of a target is affected
by the probability of existence of far-away independent targets. This
effect translates to metric-driven sensor management by producing
counterintuitive effects in the optimal actions. Though not included
in this analysis, the complete OSPA metric \cite{Vu20} also suffers
from spooky effect in optimal estimation, and entanglement of optimal
actions in sensor management.
\end{example}

\section{Conclusion\label{sec:Conclusion}}

This paper has presented two closed-form analyses for multi-target
sensor management based on metrics. In a sensor management task that
should be in principle separable for each sensor, only the GOSPA metric
preserves this property. With GOSPA-driven sensor management, we aim
to minimise localisation errors and the number of missed and false
targets. This is an intuitive objective to drive sensor actions. This
metric is also suitable to perform sensor management in large areas
due to spatial separability.

On the contrary, even in separable sensor management problems, OSPA
and UOSPA seek more complicated policies in which the optimal decision
of a sensor is influenced by far away, independent potential targets.
This paper therefore encourages the use of GOSPA for metric-driven
sensor management for usual multiple target estimation tasks. The
approach can also be extended to non-myopic sensor management with
a metric for sets of trajectories \cite{Angel20_d}, or the sum of
the multi-target metric errors in a window of future time steps.

\appendices{}

\section{\label{sec:AppendixA}}

In this appendix, we provide the proofs of Analysis I in Section \ref{sec:Analysis-I}.
In particular, we prove the cost equations (\ref{eq:c0_example1})
and (\ref{eq:c1_example1}), and the optimal action equations, see
(\ref{eq:optimal_action_analysisI_1}) and (\ref{eq:optimal_action_analysisI_2}). 

The proof of (\ref{eq:c0_example1}) and (\ref{eq:c1_example1}) requires
the following calculations: density of the measurement (\ref{eq:density_measurement}),
posterior density (\ref{eq:posterior}), optimal estimate $\hat{X}\left(Z,a\right)$
and $\mathrm{MMSGOSPA}\left(Z,a\right)$, and finally evaluation of
(\ref{eq:c_action}).

\subsection{Density of the measurement}

We calculate $p\left(Z;a\right)$ in (\ref{eq:density_measurement})
for $a=1$ and $a=0$. For $a=0$,
\begin{align}
p\left(Z;0\right) & =\delta_{\emptyset}\left(Z\right).\label{eq:p_Z0_Append1}
\end{align}

For $a=1$, we have
\begin{align}
p\left(Z;1\right) & =\int f\left(Z|X;1\right)f\left(X\right)\delta X\nonumber \\
 & =\left(1-r\right)f\left(Z|\emptyset;1\right)+rf\left(Z|\left\{ \overline{x}\right\} ;1\right)\nonumber \\
 & =\begin{cases}
1-rp^{D} & Z=\emptyset\\
rp^{D}l\left(z|\overline{x}\right) & Z=\left\{ z\right\} \\
0 & \mathrm{otherwise}
\end{cases}\label{eq:p_Z_Append1}
\end{align}
where we have used (\ref{eq:Bernoulli_density}) and the properties
of the measurement model. Note that $Z=\emptyset$ if the target does
not exist, which occurs with probability $1-r$, or the target exists
but is not detected, which occurs with probability $r\left(1-p^{D}\right)$.

\subsection{Posterior density}

We calculate $p\left(X|Z;a\right)$ in (\ref{eq:posterior}) for $a=1$
and $a=0$. For $a=0$, the measurement $Z=\emptyset$ with probability
one, and the posterior is equal to the prior
\begin{align*}
p\left(X|\emptyset;0\right) & =p\left(X\right)
\end{align*}
where $p\left(X\right)$ is given by (\ref{eq:Bernoulli_density}).
For $a=1$, we have two cases: $Z=\emptyset$ and $Z=\left\{ z\right\} $.
Applying Bayes' rule, we obtain
\begin{align*}
p\left(X|\emptyset;1\right) & =\frac{1}{1-r+r\left(1-p^{D}\right)}\\
 & \:\times\begin{cases}
1-r & X=\emptyset\\
r\left(1-p^{D}\right)\delta\left(x-\overline{x}\right) & X=\left\{ x\right\} \\
0 & \mathrm{otherwise}.
\end{cases}
\end{align*}
For $Z=\left\{ z\right\} $, we obtain
\begin{align*}
p\left(X|\left\{ z\right\} ;1\right) & =\begin{cases}
\delta\left(x-\overline{x}\right) & X=\left\{ x\right\} \\
0 & \mathrm{otherwise}.
\end{cases}
\end{align*}

\subsection{Optimal estimate for GOSPA}

The optimal estimate for GOSPA when the posterior is multi-Bernoulli
with deterministic target state was calculated in \cite{Angel19_d}.
We detect a target if its probability of existence is higher than
0.5, or misdetect it otherwise. We therefore have
\begin{align*}
\hat{X}\left(\emptyset,0\right) & =\begin{cases}
\emptyset & r<0.5\\
\left\{ \overline{x}\right\}  & \mathrm{otherwise}
\end{cases}\\
\hat{X}\left(\emptyset,1\right) & =\begin{cases}
\emptyset & \frac{1-r}{1-rp^{D}}<0.5\\
\left\{ \overline{x}\right\}  & \mathrm{otherwise}
\end{cases}\\
\hat{X}\left(\left\{ z\right\} ,1\right) & =\left\{ \overline{x}\right\} .
\end{align*}

\subsection{Action costs}

We first calculate $\mathrm{MMSGOSPA}\left(Z,a\right)$ in (\ref{eq:MMSGOSPA}),
which corresponds to substituting the above optimal estimates into
the MSGOSPA error, see \cite[Eq.(5)]{Angel19_d}, to obtain
\begin{align}
\mathrm{MMSGOSPA}\left(\emptyset,0\right) & =\frac{c^{2}}{2}t\left(r\right)\label{eq:MMSGOSPA_append1_1}\\
\mathrm{MMSGOSPA}\left(\emptyset,1\right) & =\frac{c^{2}}{2}t\left(\frac{1-r}{1-rp^{D}}\right)\nonumber \\
 & =\frac{c^{2}}{2}t\left(\frac{r\left(1-p^{D}\right)}{1-rp^{D}}\right)\\
\mathrm{MMSGOSPA}\left(\left\{ z\right\} ,1\right) & =0\label{eq:MMSGOSPA_append1_2}
\end{align}
where function $t\left(\cdot\right)$ is defined in (\ref{eq:t_function}).

We finish the proof of $c\left(0\right)$ and $c\left(1\right)$ ((\ref{eq:c0_example1})
and (\ref{eq:c1_example1})) by substituting (\ref{eq:p_Z0_Append1}),
(\ref{eq:p_Z_Append1}) and (\ref{eq:MMSGOSPA_append1_1})-(\ref{eq:MMSGOSPA_append1_2})
into (\ref{eq:c_action}).

\subsection{Optimal action}

We proceed to calculate the optimal action $\hat{a}$. Function $t\left(\cdot\right)$
is a piece-wise function, which implies that (\ref{eq:c0_example1})
and (\ref{eq:c1_example1}) are piece-wise as well. Cost $c\left(0\right)$
has two segments $r<0.5$ and $r\geq0.5$. The cost $c\left(1\right)$
has two segments
\begin{equation}
r<\frac{1}{2-p^{D}},\quad r\geq\frac{1}{2-p^{D}}.
\end{equation}
We then have four segments in $r$ for which we should compute the
minimum between $c\left(0\right)$ and $c\left(1\right)$. However,
it is met that
\begin{align}
\frac{1}{2-p^{D}} & \geq0.5
\end{align}
which implies that we actually have three regions of interest: $r<0.5$,
$0.5\leq r<\frac{1}{2-p^{D}}$ and $r\geq\frac{1}{2-p^{D}}$.

\subsubsection{Region 1}

For $r<0.5$, we have
\begin{equation}
c\left(0\right)=\frac{c^{2}}{2}r,\quad c\left(1\right)=\frac{c^{2}}{2}r\left(1-p^{D}\right)+s.
\end{equation}
We take action 1 if $c\left(1\right)<c\left(0\right)$, which implies
that
\begin{align}
r & >\frac{2s}{c^{2}p^{D}}.
\end{align}

\subsubsection{Region 2 }

For $0.5\leq r<\frac{1}{2-p^{D}}$, we have
\begin{align}
c\left(0\right) & =\frac{c^{2}}{2}\left(1-r\right)\\
c\left(1\right) & =\frac{c^{2}}{2}r\left(1-p^{D}\right)+s.
\end{align}
We take action 1 if
\begin{align}
r & <\frac{\frac{c^{2}}{2}-s}{c^{2}\left(1-\frac{1}{2}p^{D}\right)}.
\end{align}

\subsubsection{Region 3}

For $r_{1}>\frac{1}{2-p^{D}}$, we have
\begin{align}
c\left(0\right) & =\frac{c^{2}}{2}\left(1-r\right)\\
c\left(1\right) & =\frac{c^{2}}{2}\left(1-r\right)+s.
\end{align}
The costs are the same for $s=0$. We never take action 1 for $s>0$.
Putting all the results together, we finish the proof of the expression
for the optimal action in (\ref{eq:optimal_action_analysisI_1}) and
(\ref{eq:optimal_action_analysisI_2}).

\section{\label{sec:AppendixB}}

In this appendix, we prove the results of Analysis II using GOSPA
in Section \ref{sec:Analysis-II}.

\subsection{Density of the measurement\label{subsec:Append_density_measurement}}

The density of the measurement set for action $a$ is
\begin{align}
p\left(Z;a\right) & =\sum_{Z_{1}\uplus...\uplus Z_{N}=Z}\prod_{i=1}^{N}p_{i}\left(Z_{i};a_{i}\right)\label{eq:density_measurement_append}
\end{align}
\begin{align}
p_{i}\left(Z;1\right) & =\begin{cases}
1-r_{i}p^{D} & Z=\emptyset\\
r_{i}p^{D}l\left(z|\overline{x}_{i}\right) & Z=\left\{ z\right\} \\
0 & \mathrm{otherwise}
\end{cases}\label{eq:density_measurement_action1_append}
\end{align}
\begin{align}
p_{i}\left(Z;0\right) & =\delta_{\emptyset}\left(Z\right).
\end{align}
These equations extend the results for one Bernoulli, and are analogous
to the multi-Bernoulli filter prediction step \cite{Williams15b}.

\subsection{Posterior density\label{subsec:Append_Posterior-density}}

A measurement set $Z$ can be written as $Z=Z_{1}\uplus...\uplus Z_{N}$
where $Z_{i}=\emptyset$ if target located at $\overline{x}_{i}$
has not been detected and $Z_{i}=\left\{ z\right\} $ with $z\sim l\left(\cdot|\overline{x}_{i}\right)$
if the target is detected. As targets are far away and $l\left(\cdot|\overline{x}_{i}\right)$
meets A3, given $Z$, we can recover $Z_{1},...,Z_{N}$ such that
$Z=Z_{1}\uplus...\uplus Z_{N}$. 

The posterior $p\left(X|Z;a\right)$ for $Z=Z_{1}\uplus...\uplus Z_{N}$
can be written as
\begin{align}
p\left(X|Z;a\right) & =\sum_{X_{1}\uplus...\uplus X_{N}=X}\prod_{i=1}^{N}p_{i}\left(X_{i}|Z_{i};a_{i}\right)\label{eq:posterior_append}
\end{align}
where, for $a_{i}=0$,
\begin{align}
p_{i}\left(X_{i}|\emptyset;0\right) & =p\left(X\right)
\end{align}
and, for $a_{i}=1$,
\begin{align}
p_{i}\left(X_{i}|\emptyset;1\right) & =\frac{1}{1-r_{i}+r_{i}\left(1-p^{D}\right)}\nonumber \\
 & \:\times\begin{cases}
1-r_{i} & X=\emptyset\\
r_{i}\left(1-p^{D}\right)\delta\left(x-\overline{x}_{i}\right) & X=\left\{ x\right\} \\
0 & \mathrm{otherwise}.
\end{cases}
\end{align}
For $Z=\left\{ z\right\} $, we obtain
\begin{align}
p_{i}\left(X_{i}|\left\{ z\right\} ;1\right) & =\begin{cases}
\delta\left(x-\overline{x}_{i}\right) & X=\left\{ x\right\} \\
0 & \mathrm{otherwise}.
\end{cases}
\end{align}
This result can be obtained from the Poisson multi-Bernoulli mixture
update \cite{Williams15b,Angel18_b} and the multi-Bernoulli mixture
update \cite{Angel19_e}. Due to A3, only the local hypotheses that
link the $i$-th Bernoulli with $Z_{i}\neq\emptyset$ or with $Z_{i}=\emptyset$
exist.

\subsection{Optimal estimate and cost for GOSPA}

The estimate that minimises the MSGOSPA error for measurement $Z$
and action $a$ is \cite{Angel19_d}
\begin{align*}
\hat{X}\left(Z,a\right) & =\hat{X_{1}}\left(Z_{1},a_{1}\right)\cup...\cup\hat{X}_{N}\left(Z_{N},a_{N}\right)
\end{align*}
\begin{align*}
\hat{X}_{i}\left(\emptyset,0\right) & =\begin{cases}
\emptyset & r_{i}<0.5\\
\left\{ \overline{x}_{i}\right\}  & \mathrm{otherwise}
\end{cases}\\
\hat{X}_{i}\left(\emptyset,1\right) & =\begin{cases}
\emptyset & \frac{1-r_{i}}{1-r_{i}p^{D}}<0.5\\
\left\{ \overline{x}_{i}\right\}  & \mathrm{otherwise}
\end{cases}\\
\hat{X}_{i}\left(\left\{ z\right\} ,1\right) & =\left\{ \overline{x}_{i}\right\} .
\end{align*}

The MSGOSPA error is additive for each Bernoulli \cite{Angel19_d}.
Then, the cost $c\left(a\right)$ for GOSPA becomes (\ref{eq:cost_GOSPA_II_a})
and (\ref{eq:cost_GOSPA_II_c}), where $t\left(\cdot\right)$ is defined
in (\ref{eq:t_function}) and we have used the result obtained for
one Bernoulli component.

\section{\label{sec:AppendixC}}

The expressions for the optimal estimate and action cost for general
$N$ for OSPA and UOSPA are more complex than for GOSPA. We develop
the expression (\ref{eq:c_action}) to be able to obtain the optimal
action for the case $N=2$. We focus on the OSPA case, and then point
out how to extend it to UOSPA.

\subsection{Minimum mean square error}

We denote a possible (optimal) estimate of the set of targets using
existence variables $\hat{e}_{1}$ and $\hat{e}_{2}$, where $\hat{e}_{i}=1$
if the $i$-th Bernoulli component is detected and zero otherwise.
The estimated set is
\begin{align}
\hat{X} & =\left\{ \overline{x}_{i}:\hat{e}_{i}=1,\,i\in\left\{ 1,2\right\} \right\} .
\end{align}

The MSOSPA error of this estimate when the posterior existence of
the Bernoullis are $r_{1}$ and $r_{2}$ is \cite{Angel19_d}
\begin{align}
 & \mathrm{MSOSPA}\left(r_{1},r_{2},\hat{e}_{1},\hat{e}_{2}\right)\nonumber \\
 & =\begin{cases}
c^{2}\left(1-\sum_{i=1}^{2}\left[\hat{e}_{i}r_{i}\sum_{n=0}^{1}\frac{\rho_{-i}\left(n\right)}{\max\left(n+1,\hat{n}\right)}\right]\right) & \hat{n}>0\\
c^{2}\left(1-\rho\left(0\right)\right) & \hat{n}=0
\end{cases}\label{eq:MSOSPA_r12_append}
\end{align}
where $\hat{n}=\hat{e}_{1}+\hat{e}_{2}$ is the number of detected
targets, $\rho\left(\cdot\right)$ is the cardinality distribution
of the multi-Bernoulli density (\ref{eq:MB_density}) and $\rho_{-i}\left(\cdot\right)$
is the cardinality distribution of the multi-Bernoulli density without
the $i$-th Bernoulli component. 

The resulting MMSOSPA is
\begin{align*}
\mathrm{MMSOSPA}\left(r_{1},r_{2}\right) & =\min_{\hat{e}_{1},\hat{e}_{2}}\mathrm{MSOSPA}\left(r_{1},r_{2},\hat{e}_{1},\hat{e}_{2}\right)
\end{align*}
where the minimum can be computed by evaluating the four possible
$\left(\hat{e}_{1},\hat{e}_{2}\right)$.

We proceed to compute the factor in (\ref{eq:lem_optimal_action})
corresponding to OSPA for each action, which is denoted as
\begin{align}
c_{m}\left(a\right) & =\int\mathrm{MMSOSPA}\left(Z,a\right)p\left(Z;a\right)\delta Z.\label{eq:c_m_append}
\end{align}

It should be noted that with a slight abuse of notation we use $\mathrm{MMSOSPA}\left(r_{1},r_{2}\right)$
and $\mathrm{MMSOSPA}\left(Z,a\right)$ to denote the MMSOSPA as a
function of the Bernoulli existences involved in the calculation,
or as a function of $Z$ and $a$. To compute (\ref{eq:c_m_append}),
we need the density of the measurement and the posterior, which are
given by (\ref{eq:density_measurement_append}) and (\ref{eq:posterior_append}),
respectively.

\subsection{Action $a=\left[0,0\right]$}

The density of the measurement is $f\left(Z|\left[0,0\right]\right)=\delta_{\emptyset}\left(Z\right)$
and the posterior is equal to the prior. Therefore, 
\begin{align}
c_{m}\left(\left[0,0\right]\right) & =\mathrm{MMSOSPA}\left(r_{1},r_{2}\right).
\end{align}

\subsection{Action $a=\left[1,0\right]$}

In this case, set $Z_{2}=\emptyset$ with probability one. Therefore,
the integral over $Z$ in (\ref{eq:c_m_append}) is over a Bernoulli
component with density (\ref{eq:density_measurement_action1_append}).
Then, we have
\begin{align}
c_{m}\left(\left[1,0\right]\right) & =\left(1-r_{1}p^{D}\right)\mathrm{MMSOSPA}\left(\frac{r_{1}\left(1-p^{D}\right)}{1-r_{1}p^{D}},r_{2}\right)\nonumber \\
 & +r_{1}p^{D}\mathrm{MMSOSPA}\left(1,r_{2}\right).\label{eq:c_m10_append}
\end{align}
The cost for action $a=\left[0,1\right]$ is analogous to (\ref{eq:c_m10_append})
but interchanging $r_{1}$ and $r_{2}$.

\subsection{Action $a=\left[1,1\right]$}

The density of the measurement is multi-Bernoulli with probabilities
of existence $r_{1}p^{D}$ and $r_{2}p^{D}$. We integrate over the
four possible cases (detection and misdetection of the two targets)
\begin{align*}
c_{m}\left(\left[1,1\right]\right) & =\left(1-r_{1}p^{D}\right)\left(1-r_{2}p^{D}\right)\\
 & \times\mathrm{MMSOSPA}\left(\frac{r_{1}\left(1-p^{D}\right)}{1-r_{1}p^{D}},\frac{r_{2}\left(1-p^{D}\right)}{1-r_{2}p^{D}}\right)\\
 & +r_{1}p^{D}\left(1-r_{2}p^{D}\right)\mathrm{MMSOSPA}\left(1,\frac{r_{2}\left(1-p^{D}\right)}{1-r_{2}p^{D}}\right)\\
 & +r_{2}p^{D}\left(1-r_{1}p^{D}\right)\mathrm{MMSOSPA}\left(\frac{r_{1}\left(1-p^{D}\right)}{1-r_{1}p^{D}},1\right).
\end{align*}

We can now compute the optimal action using Lemma \ref{lem:Optimal_action}.
The UOSPA costs are obtained by using the MSUOSPA \cite[Eq. (6)]{Angel19_d}
instead MSOSPA in (\ref{eq:MSOSPA_r12_append}).

\bibliographystyle{IEEEtran}
\bibliography{9C__Trabajo_laptop_Mis_articulos_Fusion_2021_GOSPA_sensor_management_Referencias}

\end{document}